\documentclass[conference]{IEEEtran}
\IEEEoverridecommandlockouts
\usepackage{booktabs}
\usepackage{cite}
\usepackage{amsmath,amssymb,amsfonts}
\usepackage{algorithmic}
\usepackage{graphicx}
\usepackage{textcomp}
\usepackage{xcolor}
\def\BibTeX{{\rm B\kern-.05em{\sc i\kern-.025em b}\kern-.08em
    T\kern-.1667em\lower.7ex\hbox{E}\kern-.125emX}}
\begin{document}

\title{TOPIC IDENTIFICATION FOR SPONTANEOUS SPEECH: ENRICHING AUDIO FEATURES WITH EMBEDDED LINGUISTIC INFORMATION \\

\thanks{This work was supported by the Foundation for Aalto University Science and Technology, by awarding the Huawei doctoral student grant. We are grateful for the Academy of Finland projects funding number 337073 “FIN-CLARIN as a Collaborative Platform for Speech Processing” and 345790 "Understanding speech and scene with ears and eyes". The computational resources were provided by Aalto ScienceIT.}

}


\author{\IEEEauthorblockN{Dejan Porjazovski}
\IEEEauthorblockA{
\textit{Aalto University}\\
Espoo, Finland \\
dejan.porjazovski@aalto.fi}
\and
\IEEEauthorblockN{Tam\'as Gr\'osz}
\IEEEauthorblockA{
\textit{Aalto University}\\
Espoo, Finland \\
tamas.grosz@aalto.fi}
\and
\IEEEauthorblockN{Mikko Kurimo}
\IEEEauthorblockA{
\textit{Aalto University}\\
Espoo, Finland \\
mikko.kurimo@aalto.fi}
}

\maketitle

\begin{abstract}
Traditional topic identification solutions from audio rely on an automatic speech recognition system (ASR) to produce transcripts used as input to a text-based model. These approaches work well in high-resource scenarios, where there are sufficient data to train both components of the pipeline. However, in low-resource situations, the ASR system, even if available, produces low-quality transcripts, leading to a bad text-based classifier. Moreover, spontaneous speech containing hesitations can further degrade the performance of the ASR model. In this paper, we investigate alternatives to the standard text-only solutions by comparing audio-only and hybrid techniques of jointly utilising text and audio features. The models evaluated on spontaneous Finnish speech demonstrate that purely audio-based solutions are a viable option when ASR components are not available, while the hybrid multi-modal solutions achieve the best results.
\end{abstract}


\section{Introduction}
With the rapid increase of data generation and collection, there is a need to organise it in a meaningful way for efficient retrieval and analysis. One way to organise the data is by automatically categorising documents based on their topic. Categorising documents by topic is an active research area and automatic topic identification (topic ID) is a positive step forward in that direction. Topic ID is also beneficial in enhancing the human-computer interaction devices by improving their efficiency \cite{valentimulti}. Having such topic ID systems for high-resource languages can be taken for granted, whereas for languages with scarce resources, those systems might not be available. This lack of resources has shifted the focus of research towards alternative topic ID solutions.

To date, topic ID using text input is a well-explored area, unlike audio, where only a few solutions exist. The text-based approaches typically rely on distributed representations of words or bag-of-words for extracting meaningful information \cite{chitkara-etal-2019-topic, boulis2005text}. However, such techniques are not directly applicable to audio. 


The existing research on topic ID from audio typically explores a multi-model pipeline where an automatic speech recognition (ASR) system generates transcripts, and a separate classifier identifies the topics from the transcripts. One such pipeline system was explored in \cite{hazen2011mce}, where the authors extracted features from word-based ASR lattices and used them together with a novel minimum classification error technique. The authors in \cite{hazen2007topic}, on the other hand, focused on exploring different techniques for selecting which recognition lattices are most useful for topic ID. In \cite{kleynhans2014investigation}, the authors explored a different direction by utilising the gender information. They employed two gender-dependent acoustic models and a bi-gram language model to generate transcripts for the various text-based topic ID systems. Another study \cite{sun2019topic} utilised multiple text features. In their experiments, the words and graphemes, generated by HMM and CTC ASR systems, were used as two different inputs to a CNN model, showcasing the effectiveness of the multi-stream input. Although these pipeline systems for topic ID work well, they nevertheless rely on a separate ASR system to generate the transcripts.

Obtaining a satisfying performance with an ASR system, however, is challenging due to the difficulties of collecting enough data in low-resource or domain-specific scenarios. Thus, some research performs topic ID directly from audio, without relying on an ASR system. In \cite{kesiraju2017topic}, the authors applied an unsupervised approach of acoustic unit discovery to generate features for training an SVM system. A similar approach for acoustic unit discovery was explored in \cite{Liu2017}, in comparison to a dynamic time-warping solution of unsupervised term discovery. Other research \cite{dong2019end} explored an end-to-end approach for topic ID from audio. The acoustic unit discovery techniques differ from the proposed solutions in this study since we do not discover acoustic features but instead rely on audio embeddings.

While previous research [e.g., 4-10] performs topic ID on either text or audio, it is possible to combine those features. For example, the authors in \cite{price2020improved} combined audio and text features, showcasing the benefits of both modalities. Another approach for combining audio and textual modalities was explored in \cite{haque2019audio}. In their study, the authors presented a solution to learn audio-linguistic embeddings, where the acoustic information is modelled by reconstructing the spectrogram and the linguistic information by learning to transcribe the audio. Generally, by combining both audio and text features, a performance improvement is expected at a cost of increased complexity and computational time.

\begin{figure}[t]
  \centering
  \includegraphics[width=252pt, height=245pt]{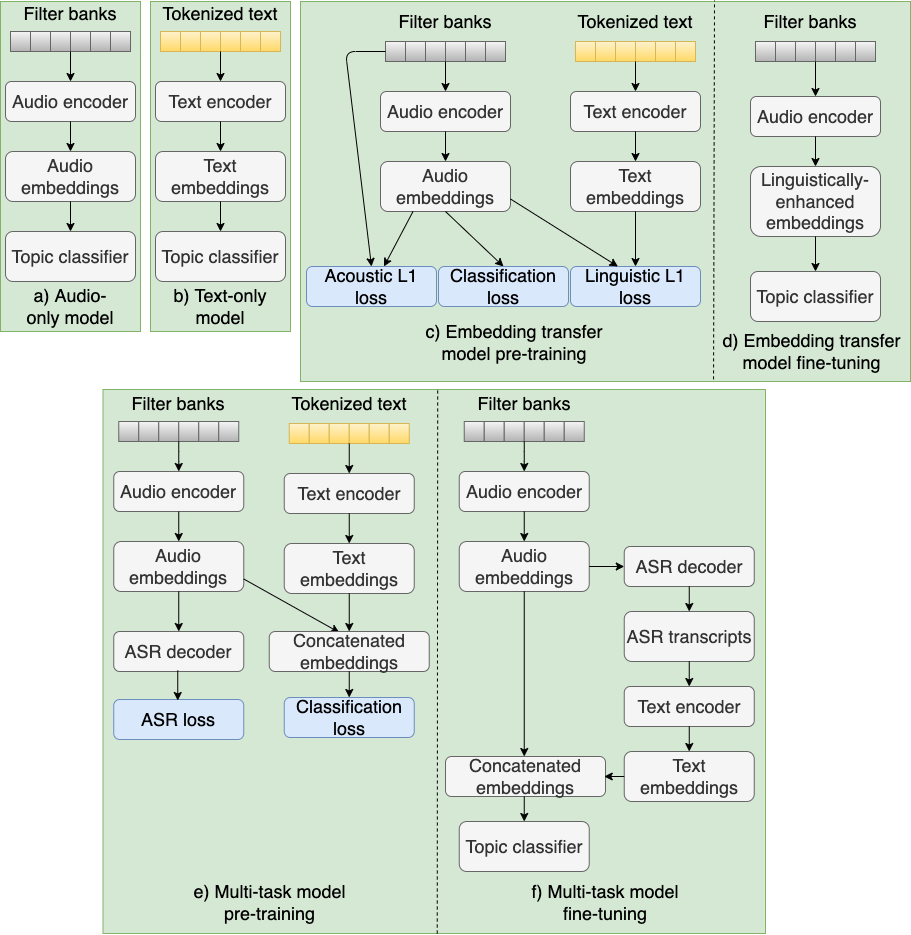}
  \caption{Overview of the topic ID systems.}
  \label{fig:topic_id_overview}
\end{figure}

Inspired by the audio-linguistic embeddings \cite{haque2019audio}, we experimented with learning linguistically-enhanced audio embeddings by reconstructing the filter banks and aligning the audio and text embedding vector spaces. This approach is different from the audio-linguistic embeddings proposed in \cite{haque2019audio} since instead of learning to transcribe the audio, we minimise the distance between the audio and text vector spaces. In addition to the audio and linguistic losses, we further include the topic ID loss as part of the learning framework. 

This work aims to investigate audio-only and hybrid techniques of jointly using audio and text-based inputs for performing topic ID from spontaneous Finnish speech and compare them to the standard text-based solutions. The text-based approaches include semi-supervised, manual transcript, and pipeline systems. The first system uses the manually and automatically (by ASR) transcribed portions of the data. The second system utilises only the manually transcribed data. The third one is a commonly-used pipeline technique for doing topic ID on the ASR-generated transcripts. As audio-only solutions, we experimented with three models. The first one uses the audio features, whereas the second one is an embedding transfer approach that learns audio embeddings, enhanced with linguistic information. The last audio-only model is a self-supervised Wav2vec2 \cite{baevski2020wav2vec} solution. To jointly utilise both the audio and text information, we experimented with two models. The first one simply concatenates the audio and text embeddings, while the second one implements multi-task training by simultaneously learning the ASR and topic ID tasks. To the best of our knowledge, this multi-task approach has not been explored for the topic ID task. The proposed audio-only and hybrid techniques are evaluated and compared against the standard text-only approaches on a colloquial Finnish dataset, reflecting a real-world scenario.  An overview of the explored topic ID systems is given in Figure \ref{fig:topic_id_overview}.

\textbf{Contributions.} In this paper, we present the effectiveness of the proposed linguistically-enhanced audio embeddings, outperforming the other audio-only solution. Furthermore, we present a multi-task approach that jointly learns the ASR and topic ID tasks as an alternative to the pipeline system.



\section{Dataset}
\begin{figure}[t]
  \centering
  \includegraphics[width=250pt, height=120pt]{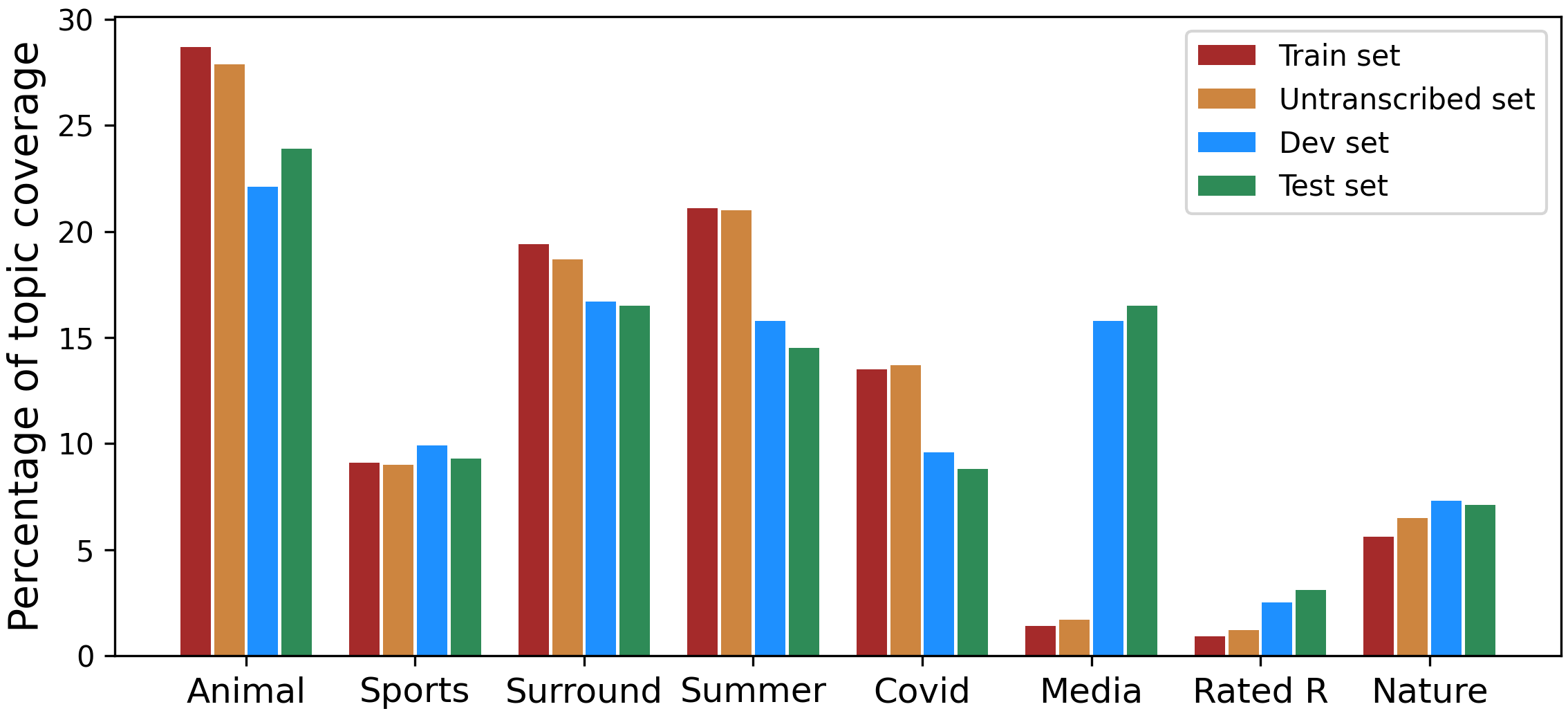}
  \caption{Topic distribution in the LP dataset.}
  \label{fig:lp_topic_dist}
\end{figure}

For conducting the experiments, we used the newly collected Lahjoita puhetta (LP) \cite{https://doi.org/10.48550/arxiv.2203.12906} conversational Finnish corpus. It is a diverse corpus with over 20.000 different speakers covering different age groups and speaking colloquial Finnish. The colloquial and spontaneous nature of the corpus allows us to evaluate the models in a real-world scenario. The length of the audio recordings varies significantly, from a few seconds to several minutes. In our experiments, we used only the samples which are up to 50 seconds long and discarded the rest. This choice reduces the computational cost that long utterances impose. The duration of the transcribed part of the dataset used in this study for the train, dev, and test splits is 408, 3, and 3 hours, respectively, whereas the untranscribed part is 405 hours (used exclusively for training). The 8 topics used in the experiments are: ”Animal friends”, ”Sports moments”, ”My surroundings”, ”Summer”, ”The cursed Covid”, ”Media skills”, ”Rated R”, and ”Nature”. In the original dataset, the ”Media skills” topic is divided into three categories, which we merged into one. The topic distribution is given in Figure \ref{fig:lp_topic_dist}.

\section{Topic identification approaches}
All the explored approaches consist of one or multiple of the following blocks: audio encoder, text encoder, and attention-based decoder. The CRDNN audio encoder uses filter banks with 40 filters as input and produces audio embeddings. The text encoder extracts word embeddings using the uncased FinBERT model \cite{virtanen2019multilingual}. The embeddings are averaged to get an utterance-level representation. The GRU decoder is used in cases where the model additionally performs the ASR task. As a decoding strategy for the ASR, we used beam search with a size of 10. During training, we optimised the negative log-likelihood. In the experiments where additionally the ASR decoder is trained, the CTC objective \cite{graves2006connectionist} is used as an auxiliary loss for the first 20 epochs with a weighting factor of 0.5. We used the SpeechBrain toolkit to conduct all the experiments and the recipes are publicly available\footnote{https://github.com/aalto-speech/Topic-identification-for-spontaneous-Finnish-speech}.

We compare our proposed audio-only and hybrid solutions to the baselines presented in \cite{https://doi.org/10.48550/arxiv.2203.12906}. The audio-only baseline model (referred to as \textit{audio BL} in the table) is a 5-layer TDNN, followed by a statistics pooling and two linear layers. The audio+text baseline (referred to as \textit{audio+text BL}) additionally utilises the text by extracting FinBERT embeddings and using them as input to a BLSTM network. The predictions are made by concatenating the audio and text embeddings.

\subsection{Text-based approaches}
Three different text-based models are used to compare the performance against the audio-only and hybrid solutions. The first one is trained on part of the data that have manually annotated transcripts. We will refer to this model as \textit{text man\_trn}. The second model combines manually and automatically generated transcripts. The automatically generated transcripts for the untranscribed portion of the data are obtained using a separate ASR system. We will call this approach \textit{text semi-supervised}. The ASR system is trained with the standard end-to-end sequence-to-sequence technique \cite{watanabe2017hybrid}, where the model outputs SentencePiece unigram subword units \cite{kudo-richardson-2018-sentencepiece}. The WER and CER on the test set are 34.63\% and 12.55\%, respectively. The reason for the high WER and CER might be the agglutinative nature of the Finnish language, which has long words. Furthermore, the colloquial Finnish does not have a fixed written form, making it hard to transcribe even by humans \cite{https://doi.org/10.48550/arxiv.2203.12906}. Lastly, we trained a \textit{text pipeline} system where all the transcripts are generated by the ASR model.

\subsection{Audio-based approaches}
The audio-based solutions rely on the audio encoder to extract meaningful acoustic embeddings. The \textit{audio-only} system embeds the utterances using the CRDNN encoder and uses those embeddings for classification (see Figure \ref{fig:topic_id_overview}a). Since we do not rely on the transcripts, the whole audio data is used, combining the transcribed and untranscribed parts.

As mentioned earlier, extracting meaningful information from audio is more difficult than from text. To bridge the gap between audio and text embeddings, we make use of the audio and text data by learning linguistically-enhanced audio embeddings. We do that by pre-training the \textit{embedding transfer} model using three objectives (see Figure \ref{fig:topic_id_overview}c). The linguistic information is modelled by minimising the L1 loss between the audio and word embeddings. Due to the different embedding dimensions, the audio embeddings are upscaled using a feed-forward layer to match the dimension of the text embeddings. To model the acoustic information, we minimise the L1 loss between the audio embeddings and the original filter bank features. The audio embeddings are downscaled using a feed-forward layer to match the dimension of the filter banks. This acoustic loss is used to get a system similar to an audio autoencoder. We opted for that to ensure the model preserves the acoustic information rather than solely adapting to the text embeddings. When computing the L1 loss between the different embeddings, we are using utterance-level representations, obtained by averaging. Lastly, the negative log-likelihood loss is used to learn the topic ID task from the audio embeddings. During pre-training, the three objective functions are interpolated with equal contribution. The negative log-likelihood loss is defined as:
\begin{equation}
    \label{eq:1}
    Loss_{nll} = -\sum^N_{i=1}log (p(y_i|a_i; \theta))
\end{equation}
where, $N$ is the number of samples, $ y_i $ is the ground truth label for the sample $ i $, $ a_i $ is the audio embedding for the sample $ i $, and $\theta$ are the model parameters.
The acoustic and linguistic L1 losses are defined as:
\begin{equation}
  Loss_{L1} = \sum_{i=1}^N |a_i - k_i|
  \label{eq2}
\end{equation}
where $k$ is either the filter bank features or the text embedding for the sample $i$. After the model is pre-trained, a new model is initialised using the pre-trained weights and trained on the whole audio data in the same way as the \textit{audio-only} system (see Figure \ref{fig:topic_id_overview}d). We put the \textit{embedding transfer} approach under the audio-based systems because during inference it uses only the audio features.

The last audio-only solution is the pre-trained multilingual Wav2vec2 \footnote{https://huggingface.co/facebook/Wav2vec2-xls-r-300m} model. We fine-tuned this model for 10 epochs on a 100-hour subset of the Lahjoita puhetta corpus. The samples were chosen so that they would reflect the class distribution of the whole dataset. We did not use the whole dataset during fine-tuning due to the computational time. Additionally, at least in terms of WER, there are no big differences observed when fine-tuning on a whole dataset, in comparison to a 10-hour subset \cite{hsu21_interspeech}.

\subsection{Hybrid audio and text-based approaches}
The hybrid systems exploit the information from both the audio and text modalities. The first solution employs a multi-task approach where the ASR and topic ID tasks are learned jointly using the transcribed portion of the data (see Figure \ref{fig:topic_id_overview}e). This way, the ASR part is optimised for the topic ID task and vice-versa. In the process, hard parameter sharing is employed \cite{ruder2017overview}, where the same audio encoder output is used in both tasks. For the topic ID task, the embeddings produced from the audio and text encoders are concatenated and passed to the classification layer. To learn the ASR task, only the audio embeddings are used. During training, both loss functions are interpolated with equal contribution. The epoch resulting in the lowest WER of 35.11\% is chosen as a starting point for the fine-tuning step. In the fine-tuning process, in addition to the transcribed portion, the untranscribed part is used by first generating transcripts using the pre-trained model. The model is fine-tuned only on the topic ID task by concatenating the audio and text modalities, using the whole training data (see Figure \ref{fig:topic_id_overview}f). In the tables, we refer to this model as \textit{multi-task}.

In the last experiment, only the transcribed portion of the data is exploited. The topic ID task is learned by concatenating the audio and text embeddings and using the output for classification. This \textit{audio+text} approach does not incorporate any pre-training and is used to show whether utilising the audio data along with the text is better than just the text.

\section{Results and discussion}
To account for the imbalanced class distribution in the Lahjoita puhetta dataset, the systems are evaluated using the micro $F_1$ and unweighted average recall (UAR) metrics. The results obtained by the text-only, audio-only, and hybrid models are presented in Table \ref{tab:lp_results}.

\begin{table}[t]
  \centering
  \caption{Micro $F_1$ and UAR scores on the dev and test sets for the text-based (rows 1-3), audio-based (rows 4-7), and hybrid (rows 8-10) models.}
  \begin{tabular}{l|l|ll|lll}
    \toprule
    \textbf{Model} & \textbf{Params} & \textbf{F1 dev} & \textbf{F1 test} & \textbf{UAR dev} & \textbf{UAR test} \\ \midrule
    Text pipeline & 7.7K & 85.99 & 80.38 & 80.30 & 75.15 \\
    Text man\_trn & 7.7K & 87.22 & 83.97 & 81.40 & 78.86\\
    Text semi-sup & 7.7K & \textbf{87.96} & \textbf{84.68} & \textbf{83.70} & \textbf{80.40}\\
    \cmidrule(lr){1-6}
    Audio BL & 4.2M & 68.28 & 65.65 & 58.34 & 58.34 \\
    Emb transfer & 21.2M & 75.67 & \textbf{76.31} & 69.32 & \textbf{72.56}\\
    Audio-only & 20.8M & 76.90 & 75.83 & 69.23 & 68.50 \\
    Wav2vec2 & 317M & \textbf{78.62} & 71.77 & \textbf{70.37}  & 67.29 \\
    \cmidrule(lr){1-6}
    Audio+text BL & 15.8M & \textbf{89.95} & 81.34 & \textbf{85.80} & 78.70 \\
    Audio+text & 20.8M & 86.48 & 85.64 & 81.69 & 79.80 \\
    Multi-task & 25M & 88.69 & \textbf{86.12} & 83.33 & \textbf{80.40} \\
    \bottomrule
  \end{tabular}
  \label{tab:lp_results}
\end{table}

In the first three rows of Table \ref{tab:lp_results}, we can see the $F_1$ and UAR results for the text-based approaches. The \textit{text pipeline}, which is the most commonly used approach in the literature, achieves an $F_1$ score of 80.38\% on the test set. Even though this model uses the ASR-generated transcripts from the whole data, it still falls behind the \textit{text man\_trn} system, trained only on the manually transcribed portion, highlighting the importance of accurate transcriptions. The benefits of the additional non-transcribed data are shown by the \textit{text semi-supervised} model, achieving the best performance of 84.68\% $F_1$ score on the test set. A similar trend can be observed by looking at the UAR metric, where the \textit{text semi-supervised} model got 80.40\%, which is a 5.25\% absolute improvement over the pipeline system. The gap between UAR and $F_1$ showcases that the models somewhat overfitted to the over-represented topics, but the overfit is not severe.

From the results produced by the audio-based approaches (rows 4-7), we can notice that the \textit{embedding transfer} model slightly outperforms the \textit{audio-only} model in terms of $F_1$ score on the test set, while in terms of the UAR metric, the difference is more substantial. On the dev set, the F1 score of the \textit{audio-only} model is better, but its UAR value is close to the one yielded by the \textit{embedding transfer}. Additionally, both models outperformed the baseline system, showcasing the benefits of the larger CRDNN model. Even though the Wav2vec2 model is pre-trained on 436K hours of data and has over 300 million parameters, it still falls behind the proposed audio-only solutions when evaluated on the test set. On the dev set, on the other hand, the Wav2vec2 model achieves the best results, possibly due to overfitting.

Next, we analyse the results achieved by combining the audio and text-based features, shown in rows 8-10. The \textit{audio+text} approach, which uses only the transcribed part of the data, achieves an $F_1$ score of 85.64\% on the test set. This is a 1.67\% absolute improvement over the \textit{text man\_trn} model that utilises only the manually-transcribed text features. Similarly, an improvement can be noticed by looking at the UAR metric. These results tell us that the model additionally benefits from the audio-based features. An improvement over the \textit{audio+text} approach is realised by the \textit{multi-task} system, which uses the manually and automatically transcribed portions of the data. In terms of WER, the \textit{multi-task} system does not suffer too much, performing only 0.48\% worse than the stand-alone ASR model. The baseline \textit{audio+text BL} model achieved the best $F_1$ and UAR results on the dev set, but the worst on the test set, indicating a severe overfit.
 
Generally, by looking at the results presented in Table \ref{tab:lp_results}, we can conclude that the audio-only approaches fall behind the text-only ones. When combining audio and text-based features, on the other hand, we can see an improvement over the commonly-used pipeline approach. The best results for the $F_1$ metric are realised by the \textit{multi-task} system, whereas for the UAR, both \textit{multi-task} and \textit{text semi-supervised} achieve equal scores. Another point worth noting is that the text-based approaches tend to perform better on the dev than on the test set when evaluated using the $F_1$ metric, indicating that they might be overfitting. Meanwhile, the models that additionally use the audio features tend to perform similarly on the dev and test sets, suggesting that the audio modality serves as a regularizer for the models.

\begin{figure}[t]
  \centering
  \includegraphics[width=\linewidth]{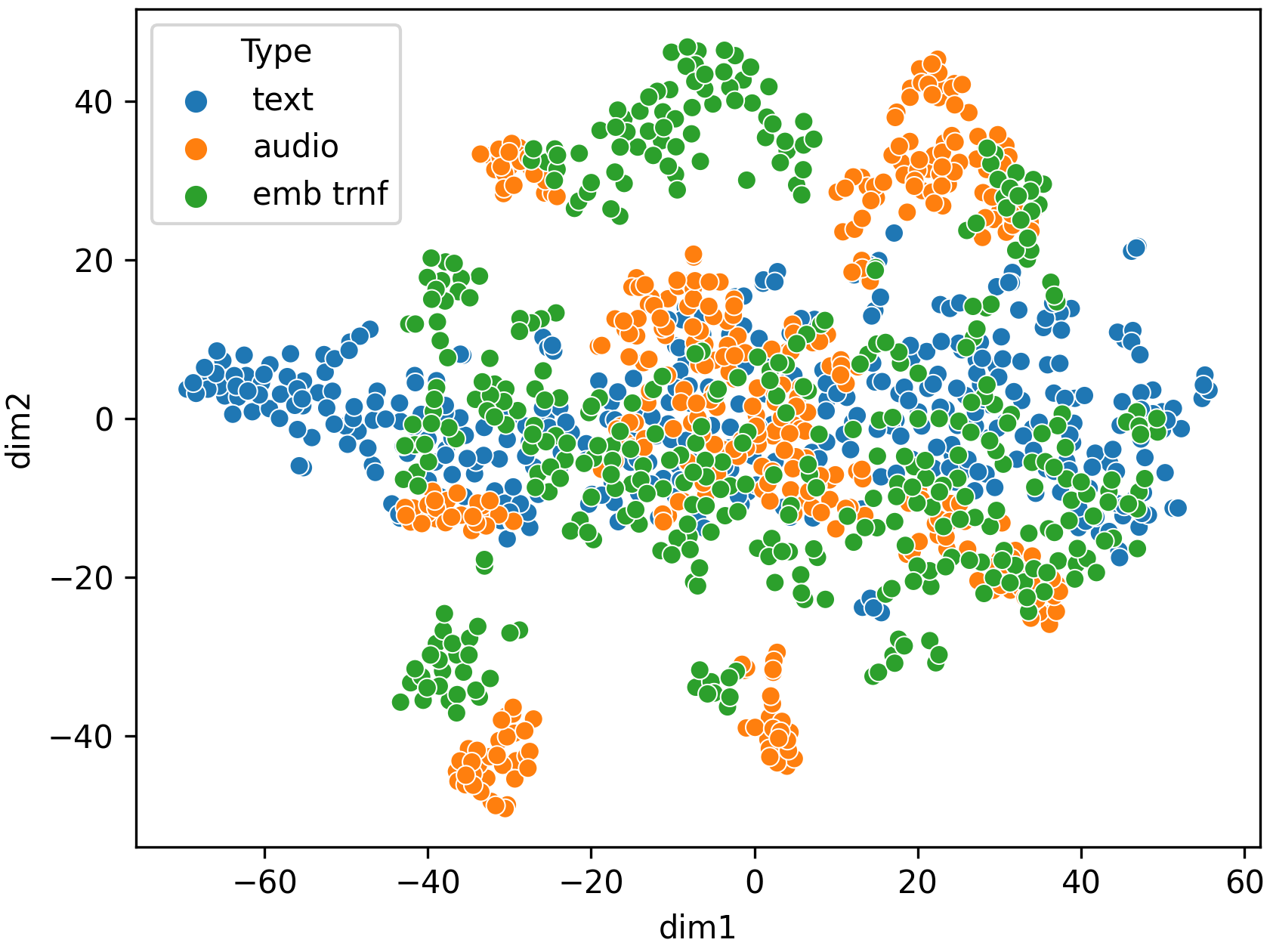}
  \caption{Embeddings produced for the test set by the \textit{audio-only}, \textit{text-only}, and \textit{embedding transfer} systems.}
  \label{fig:embeddings_tsne}
\end{figure}

\begin{table}[h]
  \centering
  \caption{Model agreement in terms of micro $F_1$ on the test set.}
  \begin{tabular}{l|lll}
    \toprule
    & {\rotatebox{0}{Audio-only}} & {\rotatebox{0}{Emb trnf}} & {\rotatebox{0}{Text semi-sup}} \\ \midrule
    Multi-task & 73.20 & 76.55 & 85.40 \\
    Audio-only & / & 72.24 & 72.72 \\
    Emb trnf & / & / & 75.59 \\
    \bottomrule
  \end{tabular}
  \label{tab:lp_model_agreement}
\end{table}

By using the input modalities in different ways, we expect that the systems would recognise the same examples with varying difficulties. To investigate how much the models agree with each other, we treated one model's predictions as true labels and compared them against the predictions of the other models. The model agreements are given in Table \ref{tab:lp_model_agreement}. These results demonstrate that the \textit{embedding transfer} solution has a higher agreement with the \textit{multi-task} and \textit{text semi-supervised} approaches, in comparison to the \textit{audio-only}, showcasing the effectiveness of modelling the linguistic information. This is further confirmed by plotting the embeddings for \textit{audio-only}, \textit{text-only}, and \textit{embedding transfer} systems, as shown in Figure \ref{fig:embeddings_tsne}. The embeddings are mapped in a 2D space using the t-SNE algorithm \cite{van2008visualizing}. From the scatter plot, we can observe that the \textit{embedding transfer} samples (green) are mostly following the distribution of the audio samples (orange), while still gaining some linguistic insights, showing that the \textit{embedding transfer} approach is capable of capturing audio and linguistic information.



\section{Conclusions}
In this paper, we experimented with text-only, audio-only, and hybrid approaches for topic ID on spontaneous Finnish speech. The experiments showed that the solutions using only audio features fall slightly behind the standard pipeline approach, but are still a viable option when the ASR system is not available. The popular Wav2vec2 model falls behind the other audio-only solutions, indicating that for the colloquial Finnish language, custom solutions trained from scratch are a better option. The \textit{embedding transfer} system, proposed in this study, was able to effectively model the audio-linguistic information, outperforming the other audio-only solutions. By combining audio and text-based modalities, we observed a significant improvement over the pipeline system and a slight improvement over the other text-based solutions. The best results are realised using the \textit{multi-task} approach of jointly learning the ASR and topic ID tasks. In the future, we plan to apply the experiments to the English language. Additionally, we plan to replace the filter banks with features extracted from a pre-trained self-supervised model.


\bibliographystyle{IEEEtran}
\bibliography{refs}

\end{document}